\documentclass[12pt]{article}
\usepackage{epsfig}     
\def\doublespace{\baselineskip=20pt}

\begin{document}

\title{\vspace*{-3cm}
\vspace*{2cm}
Clopper-Pearson Bounds from HEP Data Cuts\thanks{This research 
was partially funded by the Department of Energy under contract 
DE-FG02-97ER41022.}
      }
\author{Bernd A. Berg\\
\small Department of Physics, The Florida State University,
       Tallahassee, FL~32306, USA.\\
\small E-mail: \texttt{berg@hep.fsu.edu}\\[1.5ex]
}
\date{October 20, 2000}
\maketitle

\begin{abstract}
For the measurement of $N_s$ signals in $N$ events rigorous confidence
bounds on the true signal probability $p_{\rm exact}$ were established 
in a classical
paper by Clopper and Pearson [Biometrica 26, 404 (1934)]. Here, their
bounds are generalized to the HEP situation where cuts on the data
tag signals with probability $P_s$ and background data with likelihood
$P_b<P_s$. The Fortran program which, on input of $P_s$, $P_b$, the
number of tagged data $N^Y$ and the
total number of data $N$, returns the requested confidence bounds as
well as bounds on the entire cumulative signal distribution function,
is available on the web. In particular, the method is of interest in
connection with the statistical analysis part of the ongoing Higgs 
search at the LEP experiments.
\end{abstract}

\vfill
\noindent 
\newpage   

\doublespace

\section{Introduction}

The general theory of confidence bounds (or fiducial intervals) was 
developed by Fisher~\cite{Fi30}, Neyman and Pearson~\cite{NePe33}. 
We consider a particular problem which is of interests when cuts 
are used to analyze high energy physics data. Typically, a neural
network or some other method of performing the cuts results in
probabilities (efficiencies) to tag signals more likely than background
events. For instance by means of Monte Carlo (MC) simulations, these
probabilities can normally be calculated. Let $P_s$ be the probability
to tag a signal and $P_b$ be the likelihood to tag a background event,
$0<P_b<P_s<1$. Out of a total number of $N$ data one gets in this way
\begin{equation} \label{NY}
 N^Y ~~{\rm tagged\ data.}
\end{equation}
It is easy to find from this the mean expectation for the signal
probability. Assume that there are $N_s$ signals and $N_b$ background
events in the data. Then we have
$$ N = N_s + N_b ~~~{\rm and}~~~ N^Y=P_s\,N_s+P_b\,N_b $$
and these two equations solve for
\begin{equation} \label{P_mean}
 p_{\rm mean} = {N_s\over N} = {N^Y-P_b\,N\over N\,(P_s-P_b)}\ .
\end{equation}
The question is, what are the implied confidence limits on the signal
probability?

The special case $P_s=1$ and $P_b=0$ (sure signal detection) has
been treated by Clopper and Pearson~\cite{ClPe34} in 1934. After
briefly reviewing their approach in the next section, we derive and
illustrate the general case in section~3. This is, in part, based on
Ref.\cite{BeRi97}. 
In particular, the
method is valid when the number of tagged data is small and returns
the probability $P_0$ for the case that there is no signal, i.e. that 
the exact signal probability is $p_{\rm exact}=0$. 
This is of interest for the statistical
analysis of the ongoing Higgs search at LEP~\cite{LEP}. Discovery
of the Higgs particle on the $5\sigma$ level would mean
$P_0\le 0.287\,10^{-6}$.
Subsection~\ref{Fortran} explains the use of the
corresponding Fortran programs and how to download them from the web.
Conclusions follow in the final section~\ref{Conclusions}.

\section{The Clopper--Pearson Confidence Limits}

\begin{figure}[b] 
\centerline{\psfig{figure=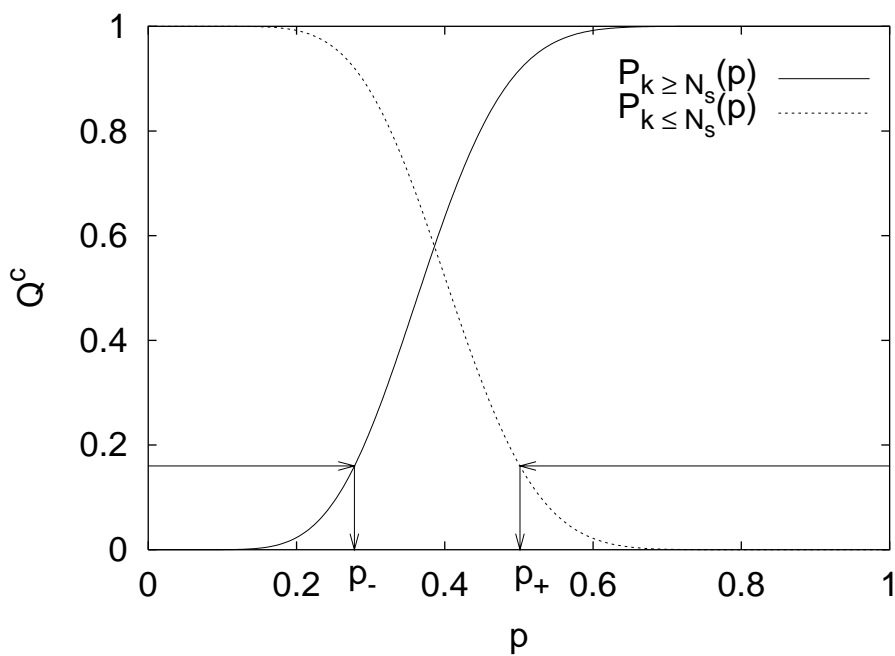}}
\caption{The probability functions $P_{k\ge N_S}(p)$ to observe 
$k\ge N_s$ signals in $N$ events (\ref{Plarge}) and $P_{k\ge N_S}(p)$ 
to observe $k\le N_s$ signals in $N$ events (\ref{Psmall}) are depicted 
for $N=26$ and $N_s=10$. Symmetric 68\% confidence bounds ala
Clopper-Pearson are also indicated. \label{fig_cp1} }
\end{figure}

Let $p$ be the likelihood that a data point is a signal. For $N$
measurements the probability to observe $k$ signals is given by the
binomial coefficient
\begin{equation} \label{binomial}
b_N(k,p) = \left(\!\!\begin{array}{c} N\\ k\end{array}\!\! \right) 
p^{k}\,q^{N-k} = {N!\over k!\, (N-k)!}\ p^{k}\, q^{N-k} 
~~{\rm with}~~ q=p-1\ .
\end{equation}
The probability to observe $k\ge N_s$ signals is given by 
\begin{equation} \label{Plarge}
  P_{k\ge N_s}(p) = \sum_{k=N_S}^{N} b_N(k,p)
\end{equation}
and the probability to observe $k\le N_s$ signals by 
\begin{equation} \label{Psmall}
  P_{k\le N_s}(p) = \sum_{k=0}^{N} b_N(k,p)\ .
\end{equation}
For $N=26$ and $N_s=10$ the functional forms of $P_{k\ge N_s}(p)$ and 
$P_{k\le N_s}(p)$ are depicted in figure~\ref{fig_cp1}.

Assume that $N_s$ signals are found in $N$ measurements and that a 
probability $Q^c<0.5$ (typical values are $Q^c=0.16$ or $Q^c=0.025$)
is given.
We can solve equation (\ref{Plarge}) for $P_{k\ge N_s}(p_-)=Q^c$ and
$p_-$ is a lower bound  on the true signal probability $p_{\rm exact}$, 
such that the likelihood to find $k\ge N_s$ signals in $N$ measurement
is smaller than $Q^c$ for every $p_{\rm exact}<p_-$. Correspondingly, 
we can solve 
equation (\ref{Psmall}) for $P_{k\le N_s}(p_+)=Q^c$ and $p_+$ is an 
upper bound on the true signal probability $p_{\rm exact}$, such that 
the likelihood to find $k\le N_s$ signals in $N$ measurement is smaller 
than $Q^c$ for every $p_{\rm exact}>p_+$. Together, this combines into 
the Clopper--Pearson bounds: The probability to find the true signal
probability in the range
\begin{equation} \label{CP_bounds}
 p_- \le p_{\rm exact} \le p_+ ~~~{\rm is\ larger\ or\ equal\ to}
 ~~~ P^c=1-2Q^c\ . 
\end{equation}
In more details the meaning of the 
inequality is discussed in~\cite{BeRi97}. For $P^c=0.68$ ($Q^c=0.16$)
the $p^{\pm}$ values are indicated in figure~\ref{fig_cp1}. 
Approximately, this range corresponds to the confidence of a $1\sigma$
error bar. Similarly the confidence range corresponding to a $2\sigma$
error bar, etc., can be found.

\section{Confidence Limits from Data Cuts}

We are interested in the situation where signal and background data
can no longer be distinguished unambiguously. Instead, a neural network 
or other device yields statistical information by tagging signals with 
efficiency $P_s$ and background data with probability $P_b$, as 
discussed in the introduction.

Applying the cuts to all $N$ data results in $N^Y$ tagged data
($0\le N^Y\le N$), composed of $N^Y=N^Y_s+N^Y_b$, where $N^Y_s$ is the 
number of tagged signals and $N^Y_b$ is the number of tagged background 
data. Of course, the values for $N^Y_s$ and $N^Y_b$ are not known. 
Our task is to determine confidence limits for the signal probability
$p$ from the sole knowledge of $N^Y$. We proceed by writing down the
probability density of $N^Y$ for given $p$ and, subsequently,
generalizing the Clopper-Pearson method.

\begin{figure}[b] 
\centerline{\psfig{figure=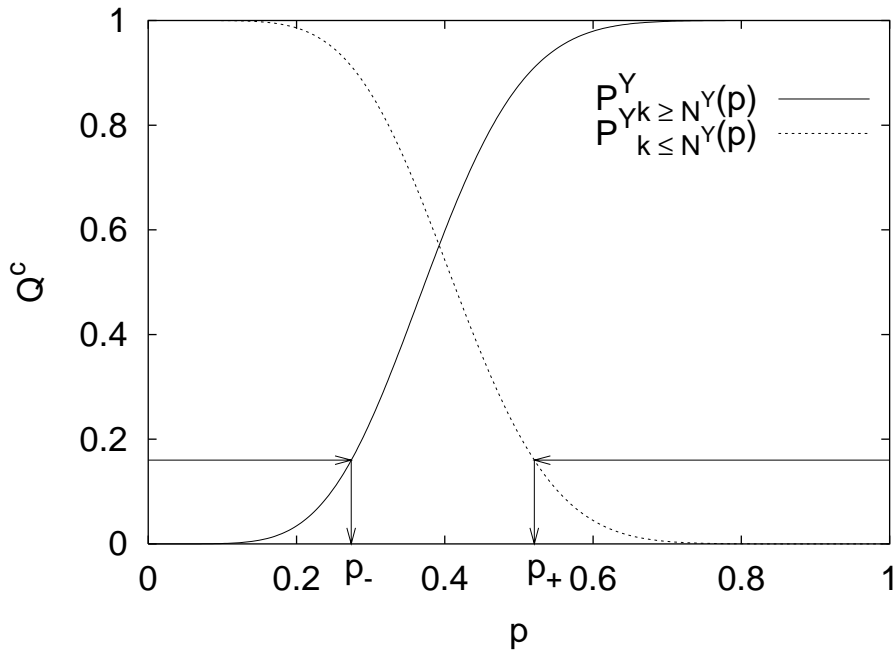}}
\caption{The probability functions $P^Y_{k\ge N_Y}(p)$ to find $k\ge N^Y$
tags in $N$ events and $P^Y_{k\le N^Y}(p)$ to find $k\le N^Y$ tags in
$N$ events (\ref{PY_l_s}) are depicted for $N=35$, $N^Y=12$, $P_s=0.8$
and $P_b=0.05$. Symmetric 68\% confidence bounds, found for $p_-=0.274$
and $p_+=0.521$, are also indicated.  \label{fig_cp2} }
\end{figure}

First, assume fixed $N_s$. The probability densities of $N^Y_s$ and 
$N^Y_b$ are binomial and thus the probability density for $N^Y$
is given by the convolution
\begin{equation} \label{PYS}
P^Y(N^Y|N_s) = \sum_{N^Y_s+N^Y_b=N^Y} b_{N_s}(N^Y_s,P_s)\, 
b_{N_b}(N^Y_b,P_b),\ N_b=N-N_s ~.  
\end{equation}
Proof: For a signal event the probability to be tagged is $P_s$, so
$b_{N_s}(N^Y_s,P_s)$ is the probability to tag $N^Y_s$ out of the
$N_s$ signals. Similarly, the probability for a background event to
become tagged is $P_b$ and $b_{N_b}(N^Y_b,P_b)$ is the probability
to tag $N^Y_b$ of the $N_b=N-N_s$ background events. As these two
probabilities are independent, the likelihood that precisely $N^Y_s$ 
of the signals and $N^Y_b$ of the background events are tagged becomes
the product $b_{N_s}(N^Y_s,P_s)\, b_{N_b}(N^Y_b,P_b)$. Summing over
all possibilities which add up to $N^Y_s+N^Y_b=N^Y$ gives the result.

Summing over $N_s$ in (\ref{PYS}) removes the constraint of fixed $N_s$
and, with $N$, $p$ fixed, the probability to tag $k$ events becomes
\begin{equation} \label{PYp}
b^Y_N(k,p) = \sum_{N_s=0}^N b_N(N_s,p)\, P^Y(k|N_s)~.
\end{equation}
Fourier transformation of the convolution~(\ref{PYS}) allows for
an efficient numerical calculation of the $P(k|N_s)$ coefficients.
In analogy with equations~(\ref{Plarge}) and~(\ref{Psmall}) we find
the probabilities to tag $k\ge N^Y$ and $k\le N^Y$ events to be 
\begin{equation} \label{PY_l_s}
  P^Y_{k\ge N^Y}(p) = \sum_{k=N^Y}^N b^Y_N(k,p) ~~{\rm and}~~
  P^Y_{k\le N^Y}(p) = \sum_{k=0}^{N^Y} b^Y_N(k,p)\ .
\end{equation}
For $N=35$, $N^Y=12$, $P_s=0.8$ and $P_b=0.05$ the 
functions~\cite{eval} $P^Y_{k\ge N^Y}(p)$ and $P^Y_{k\le N^Y}(p)$ 
are depicted in figure~\ref{fig_cp2}.
The 68\% confidence range (\ref{CP_bounds}) is 
also indicated in figure~\ref{fig_cp2}, where the bound values 
$p_{\pm}$ are now defined as solutions of the equations
\begin{equation} \label{NY_bounds}
 Q^c = P^Y_{k\ge N^Y}(p_-) ~~{\rm and}~~ Q^c = P^Y_{k\le N^Y}(p_+)\ . 
\end{equation}
The range $[p_-,p_+]$, obtained with $Q^c=0.16$, guarantees the standard 
one error bar confidence probability of 68\% for every true signal
probability $p_{\rm exact}$. For almost all values the actual confidence 
will be better. However, the bounds cannot be improved without violating
the requested confidence probability for the case that $p_{\rm exact}$ 
happens to agree with either $p_-$ or $p_+$. In the same way, bounds 
calculated with $Q^c=0.023$ ensure the standard two error bar 
confidence level of 95.4\% or better, and so on.

\subsection{Data Sets with Few Signals}

\begin{figure}[b] 
\centerline{\psfig{figure=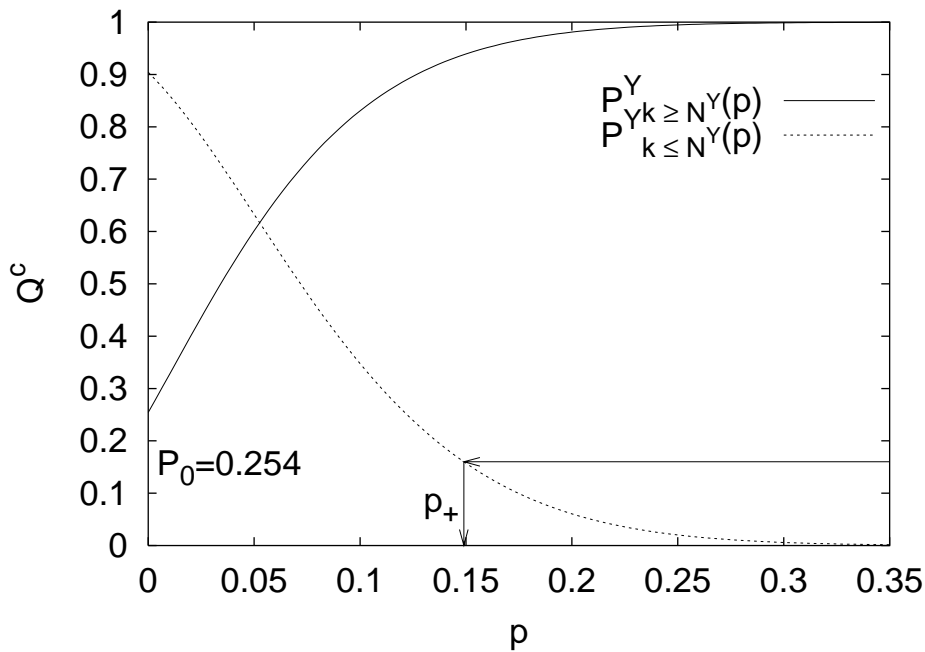}}
\caption{The same functions as in figure~\ref{fig_cp2} are depicted, 
but for $N^Y=3$ instead of $N^Y=12$ tags. The lower 68\% confidence 
bound does not exist anymore, instead the likelihood for 
$p_{\rm exact}=0$ has become $P_0=0.254$. The upper 68\% confidence 
bound is found at $p_+=0.149$.  \label{fig_cp3} }
\end{figure}

As outlined,
data sets with few signals are of of particular interest in high energy 
physics. Let us replace $N^Y=12$ of figure~\ref{fig_cp2} by $N^Y=3$. 
The resulting graph is depicted in figure~\ref{fig_cp3}. From the
$P^Y_{k\ge N^Y}(p)$ curve we read off the finite probability $P_0=0.254$
for the likelihood that the true signal probability $p_{\rm exact}=0$
generates $k\ge N_y$ tags. Due to this
probability the lower 68\% confidence bound $p_-$ disappears, whereas
the upper $p_+$ bound does still exist. In passing let us note that
for the data of figure~\ref{fig_cp2} we have $P_0=0.69\,10^{-7}$, i.e. 
there $p_{\rm exact}=0$ is ruled out on the $5\sigma$ level.

\subsection{Signal Probability Distributions}

\begin{figure}[b] 
\centerline{\psfig{figure=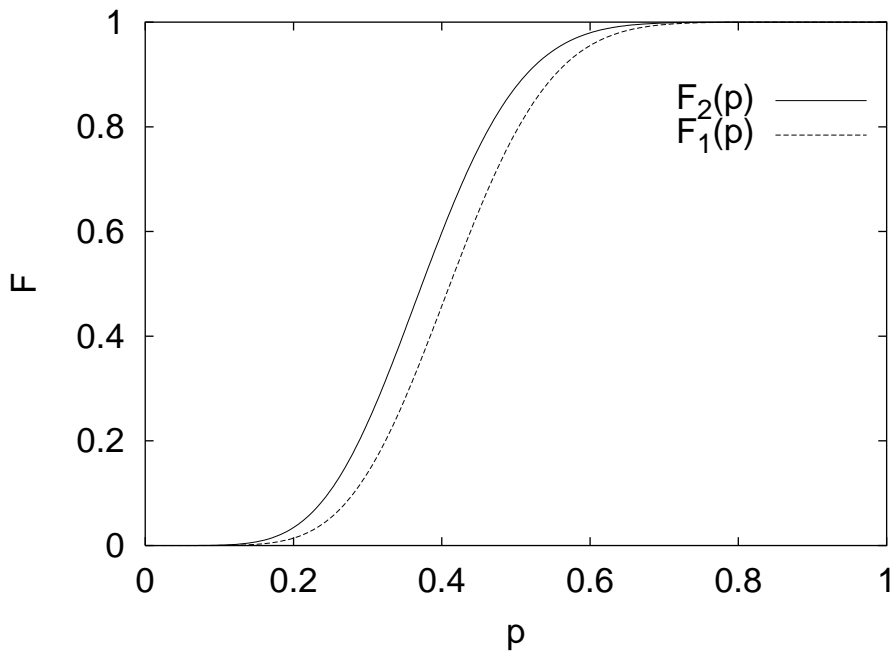}}
\caption{Upper and lower bounds, $F_2(p)$ and $F_1(p)$, for the
cumulative signal distribution function $F(p)$. The values used
for $N$, $N^Y$, $P_s$ and $P_b$ are identical with those of
figure~\ref{fig_cp2}.  \label{fig_cp4} }
\end{figure}

To avoid frequentist objections, the cumulative signal distribution
function $F(p)$ is, in the opinion of the author, best defined as
the {\it expectation of the researcher} to find the true signal
probability $p_{\rm exact}$ in the range $0\le p_{\rm exact}\le p$.
Our approach allows to estimate upper and lower bounds for the 
cumulative signal distribution function 
\begin{equation} \label{Fdefinition}
 F(p) = \int_{-\infty}^p dp'\, f(p') ~~~{\rm where}~~ f(p') 
 ~~{\rm is\ the\ signal\ probability\ density.}
\end{equation}
Equation~(\ref{NY_bounds}) implies
\begin{equation} \label{Fpbounds}
 F_1(p) = 1 - P^Y_{k\le N^Y}(p) = \sum_{k=N^Y+1}^N b^Y_N(k,p) \ \le\ 
F(p)\ \le\ F_2(p) = P^Y_{k\ge N^Y}(p) = \sum_{k=N^Y}^N b^Y(k,p)
\end{equation}
and note that $F(0)=F_2(0)$.
Figure~\ref{fig_cp4} shows the result for the same data which were
used in figure~\ref{fig_cp2}. Any reasonable Bayesian estimates
(which involves additional {\it a-priori} assumptions) should give a 
function $F(p)$ which is sandwiched between $F_1(p)$ and $F_2(p)$.

\begin{figure}[b] 
\centerline{\psfig{figure=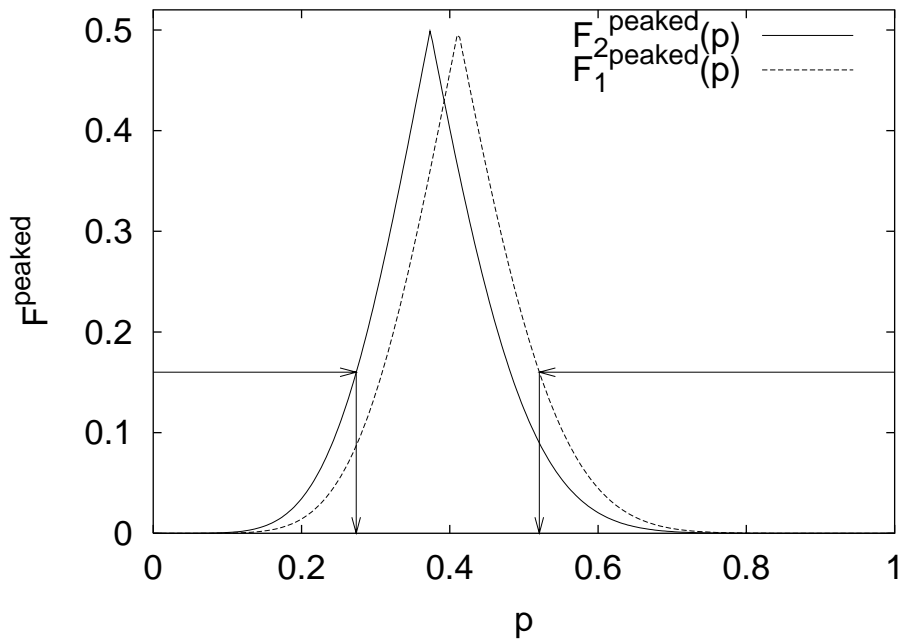}}
\caption{Peaked distribution functions $F_2^{\rm peaked}(p)$ and
$F_1^{\rm peaked}(p)$ as defined in equation~(\ref{Fpeaked}).
The values used for $N$, $N^Y$, $P_s$ and $P_b$ are identical 
with those of figures~\ref{fig_cp2} and~\ref{fig_cp4}.
\label{fig_cp5} }
\end{figure}

It is instructive to define peaked distribution
functions~\cite{Be_book} by
\begin{equation} \label{Fpeaked}
 F_i^{\rm peaked} (p) = \cases{ F_i(p) ~~{\rm for}~~ F_i(p)\le 0.5
~~{\rm and} \cr 1-F_i(p) ~~{\rm for}~~ F_i(p)\ge 0.5,~~~~~~~(i=1,2)\ .}
\end{equation}
Using the same data as in figure~\ref{fig_cp4}, $F_1^{\rm peaked}(p)$
and $F_2^{\rm peaked}(p)$ are depicted in figure~\ref{fig_cp5}. The 
advantages of using peaked distribution functions instead of
conventional cumulative distribution functions are: 
\begin{enumerate}
\item The ordinate becomes enlarged by a factor of two. 
\item The estimated medians are located at the peaks and the 
probability content of the distribution is instructively displayed.
\item Bounds like those of figure~\ref{fig_cp1} are easily read off.
\end{enumerate}

\begin{figure}[b] 
\centerline{\psfig{figure=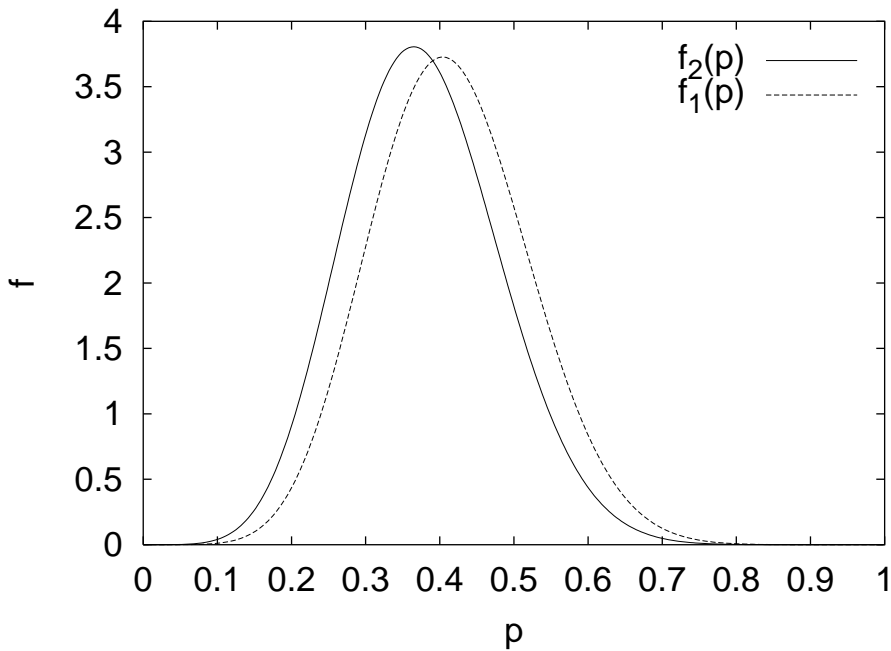}}
\caption{Probability densities $f_i(p)$ corresponding
(\ref{prob_densities}) to the cumulative distribution function of
figure~\ref{fig_cp4}. The values used for $N$, $N^Y$, $P_s$ and 
$P_b$ are identical with those of figure~\ref{fig_cp4}.
\label{fig_cp6} }
\end{figure}

The probability densities corresponding to the cumulative
distribution functions~(\ref{Fpbounds}) are the derivatives of the 
$F_i(p)$ with respect to $p$
\begin{equation} \label{prob_densities}
 f_i(p) = {d F_i(p)\over d\,p}\, ~~~(i=1,2)\ .
\end{equation}
Their numerical calculation is straightforward when analytical 
expressions for the derivatives of the binomial coefficients in 
equation~(\ref{PYp}) are used. Figure~\ref{fig_cp6} exhibits the 
results for $f_1(p)$ and $f_2(p)$ corresponding to $F_1(p)$ and
$F_2(p)$ of figure~\ref{fig_cp4}. At $p=0$ the probability 
densities have $\delta$-function contributions
\begin{equation} \label{prob_singular}
 f_i (p)\ =\ F_i(0)\, \delta (p) + ...\ ,~~~ (i=1,2)\ .
\end{equation}
In case of figure~\ref{fig_cp6} the $F_i(0)$ coefficients are
practically zero. However, for the probability densities corresponding
to figure~\ref{fig_cp3} there would be a substantial contribution:
$F_1(0)=0.096$ and $F_2(0)=0.254$ in that case.

\subsection{The Fortran Code} \label{Fortran}

The Fortran code which produces the illustrations of this paper
is available on the web. Start at the author's homepage~\cite{www}
{\tt www.hep.fsu.edu/\~\,}$\!${\tt berg} and follow the {\tt research} 
and from 
there the {\tt Clopper-Pearson} hyperlink. Load down all files of the
{\tt Fortran~Programs} subdirectory into an empty directory. Any
standard Fortran~77 compiler should then be able to compile the
{\tt cp1.f}, {\tt cp2.f} and {\tt cp3.f} programs. Be aware that
the program files include (via {\tt include} Fortran statements)
some of the other files you downloaded. Running one of the
programs produces a data files with the name of that program and
a {\tt .d} extension. Subsequently, gnuplot users can produce the
graphical presentations of this paper by using the {\tt *.plt}
driver files, as listed in the following.
\begin{eqnarray} \nonumber
 {\tt program}~~~~~~~~~~ & {\tt generates\ file} &
              ~~~~~~~~~~{\tt use\ with\ gnuplot\ driver}  \\ \nonumber
   {\tt cp1.f}~~~~~~~~~~ & {\tt cp1.d} &
              ~~~~~~~~~~{\tt cp1.plt} \\ \nonumber
   {\tt cp2.f}~~~~~~~~~~ & {\tt cp2.d} &
              ~~~~~~~~~~{\tt cp2.plt,\ cp4.plt,\ cp5.plt, cp6.plt}\\
   {\tt cp3.f}~~~~~~~~~~ & {\tt cp3.d} &
              ~~~~~~~~~~{\tt cp3.plt}     \nonumber
\end{eqnarray}
Here, the gnuplot file number corresponds to the figure number of this
paper. To get encapsulated postscript files, the comment signs in 
front of the first two rows of each gnuplot file have to be eliminated.

\section{Conclusions} \label{Conclusions}
We have calculated confidence limits, and corresponding limits of the 
entire cumulative distribution function, for an unknown true signal 
likelihood $p_{\rm exact}$. The only input used are the efficiencies 
$P_s$ for tagging signals, the probabilities $P_b$ for tagging 
background events, the number $N^Y$ of tagged data and the total
number of data $N$.
In particular, the method allows to deal with the situation where 
only few signals occur and yields then a finite probability for
the likelihood that $k\ge N^Y$ tags are observed if the true signal
probability is $p_{\rm exact}=0$. In real life the probabilities
$P_s$ and $P_b$ are most likely estimators by themselves, i.e. 
quantities with error bars. This causes no major problem, one just 
has to apply our confidence calculations to an appropriate sample
and to average over the results.

\clearpage

\begin{thebibliography}{12}

\bibitem{Fi30} R.A. Fisher, Proc. Camb. Phil. Soc. 26, 528 (1930); 
Proc. Roy. Soc. A139, 343 (1933).

\bibitem{NePe33} J. Neyman and E.S. Pearson, Phil. Trans. Roy. Soc.
A231, 289 (1933).

\bibitem{ClPe34} C.J. Clopper and E.S. Pearson, Biometrika 26, 404
(1934). For a textbook treatment see S. Brandt, {\it Statistical and
Computational Methods in Data Analysis} (North-Holland, 1983). 

\bibitem{BeRi97} B.A. Berg and J. Riedler, Comp. Phys. Commun.
107, 39 (1997).

\bibitem{LEP} {\tt http://alephwww.cern.ch/ALPUB/seminar/wds/}

\bibitem{eval} For the numerical evaluation of equations~(\ref{PY_l_s}) 
use $\sum_{k=0}^N b^Y_N(k,p)=1$ together with the partial sums from 
either $\sum_{k=N^Y}^N b^Y_N(k,p)$ or $\sum_{k=0}^{N^Y} b^Y_N(k,p)$, 
but not both. Normally $N^Y<N/2$ and $\sum_{k=0}^{N^Y} b^Y_N(k,p)$ 
will be used.

\bibitem{Be_book} B.A. Berg, {\it Introduction to Monte Carlo
Simulations and Their Statistical Analysis}, in preparation.   

\bibitem{www} The address of the authors homepage and its tree
structure are expected to be stable, whereas the absolute address
where the programs are located is likely to change.

\end{thebibliography}
\end{document}